# Electromigration Response of Microjoints in 3DIC Packaging Systems

Vahid Attari [a], Thien Duong [a], Raymundo Arroyave [a,b]
[a] Department of Materials Science and Engineering, Texas A&M University, College Station, TX, USA
[b] Department of Mechanical Engineering, Texas A&M University, College Station, TX, USA
email: attari.v@tamu.edu

## Abstract

In multilevel 3D integrated packaging, three major microstructures are viable due to the application of low volume of solders in different sizes, and/or processing conditions. Thermodynamics and kinetics of binary compounds in Cu/Sn/Cu low volume interconnection is taken into account. We show that current crowding effects can induce a driving force to cause excess vacancies saturate and ultimately cluster in the form of microvoids. A kinetic analysis is performed for electromigration mediated intermetallic growth using multi-phase-field approach. Faster growth of intermetallic compounds (IMCs) in anode layer in the expense of shrinkage of cathode IMC layer in shown. This work paves the road for computationally study the ductile failure due to formation of microvoids in low volume solder interconnects in 3DICs.

## Introduction

Solid-Liquid Interdiffusion (SLID[1]) bonding process [1,2] in Three Dimensional Integrated Circuits (3DICs) plays a crucial role in forming ultimate shape and morphology of microjoints located in between Through Silicon Vias (TSVs); and multi-phase-field modeling has proven to be a promising approach to study this process in Cu/Sn/Cu interconnections [3–5]. It is experimentally shown that electromigration (EM) induced electron wind force and the subsequent mass flux affects the diffusion in the form of atomic advection and enhances the reaction kinetics in the direction of electric field, resulting in a new dominant kinetic mechanism during operation of these devices [3,6–11]. In other words, EM changes the intermetallic compound (IMC) growth trend in the multi-layered interconnection systems. We now know that with higher current densities, the IMCs in the anode side grow faster than the IMCs in the cathode side [3,6,9,12], and the path to failure can be amplified further by current-induced temperature localization. In this respect, the study by Hu et al. [13] shows the extensive dissolution of Cu in the cathode layer and failure of the flip-chip bump only after 95 minutes, under $2.5 \times 10^8\ Am^{-2}$, while Joule heating effect also causes a temperature increase up to 157℃ at the backside of the chip.

3DIC interconnect architectures are also subject to considerable performance issues due to EM and much work remains to be done to further improving the technology. Physics-based predictive and quantitative models have the potential to greatly enable the development of 3DIC low-volume interconnect (LVI) technologies. This study addresses the need by putting forward a framework (Refer to Figure 1) that has its main objective as the development of a high-fidelity multi-physics microstructure-sensitive model to simulate mass, energy and charge transport across these complex hierarchical architectures. The model is capable of predicting the onset of damage, nucleation of microvoids and therefore loss of electric conductivity due to formation and coalescence of EM-induced microvoids. The major result of this work is a fully-coupled multi-physics software suite capable of simulating not only the formation of LVI joints but their evolution under current-stressing conditions. The proposed software tool will in turn be used to develop process and damage maps that can assist in the further development of LVI 3DIC technologies.

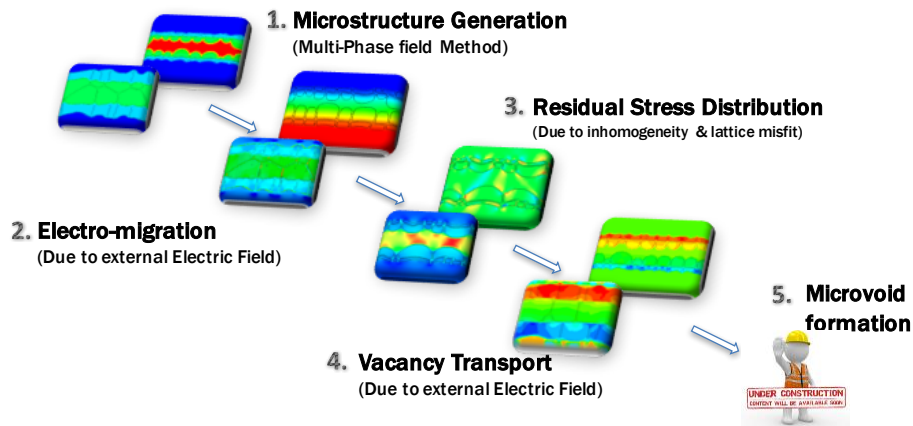

*Figure 1 The evolution of coupled multi-phase-field model over the past three years [3,5].*

---
[1] SLID is also known as Transient Liquid Phase Bonding (TLPB).

## Methods

**Multi-phase-field Model**

A multi-phase-field formalism is integrated to study the evolution of microstructure at isobaric and isothermal states. The components of the material system evolve based on the variational principles of total free energy of the material. A set of non-conserved ($\phi$) and conserved variables (c) describe the components of the microstructure, where the non-conserved variables define the spatial fraction of available phases over the domain ($\Omega$) and the conserved variables define the phase compositions. We start from a general model of total free energy of a chemically heterogeneous system that involves interfacial, bulk and electrical, and elastic interactions [14]:

$$F_{total} = \int_\Omega f^{int} + f^{bulk} + f^{elec} + f^{elas} \quad (1)$$

where the four contributing factors are respectively formulated as:

$$f^{int} = \sum_{\beta>\alpha}\sum_\alpha \frac{4\sigma_{\alpha\beta}}{w_{\alpha\beta}}\left[-\frac{w_{\alpha\beta}^2}{\pi^2}\nabla\phi_\alpha \cdot \nabla\phi_\beta + |\phi_\alpha\phi_\beta|\right] \quad (2)$$

$$f^{bulk} = \sum_\alpha \phi_\alpha f_\alpha^0(c_\alpha) \quad (3)$$

$$f^{elec} = \sum_\alpha \phi_\alpha f_\alpha^{elec}(c_\alpha) \quad (4)$$

$$f^{elas} = \frac{1}{2}\left\{\sum_\alpha \phi_\alpha (\varepsilon_\alpha^{ij} - \varepsilon_\alpha^{*ij})C_\alpha^{ijkl}(\varepsilon_\alpha^{kl} - \varepsilon_\alpha^{*kl})\right\} \quad (5)$$

where $\sigma_{\alpha\beta}$ is the interface energy, and $w_{\alpha\beta}$ is the interface width. The parabolic double-obstacle potential $|\phi_\alpha\phi_\beta|$ is defined in the interfacial region only, where $0 < \phi_{\alpha/\beta} < 1$. $f_\alpha^0$ is the homogenous free energy of phase "$\alpha$" with composition $c_\alpha$ where c is the molar fraction of Sn. $f_\alpha^{elec}$ is the energy in phase $\alpha$ due to the imposed electric field. $\varepsilon_\alpha^{ij}$, $\varepsilon_\alpha^{*ij}$, and $C_\alpha^{ijkl}$ are the elastic strain, eigenstrain, and elastic modulus of phase $\alpha$, respectively. The chemical free energies are obtained based on the study of Shim et al. [15]. The evolution of the field parameter and composition field is governed by:

$$\frac{\partial \phi_\alpha}{\partial t} = -\sum_{\alpha\neq\beta}\frac{M_{\alpha\beta}}{N_p}\left[\frac{\delta F_{total}}{\delta \phi_\alpha} - \frac{\delta F_{total}}{\delta \phi_\beta}\right] \quad (6)$$

$$\frac{\partial c}{\partial t} = \nabla \cdot \left[D(\vec{\phi})\sum_\alpha \phi_\alpha \nabla c_\alpha - \frac{D(\vec{\phi})}{K_BT}\sum_\alpha c_\alpha\phi_\alpha \cdot eZ^*(\phi_\alpha)\psi\right] \quad (7)$$

$$\nabla \cdot [\kappa(c,\vec{\phi}) \cdot \nabla \psi] = 0 \quad (8)$$

$$\frac{\partial \sigma_{ij}^{elas}}{\partial r_j} = 0 \quad (9)$$

$$\frac{\partial c_v}{\partial t} = \nabla \cdot \left[D_v^1(\vec{\phi})\nabla c_v - \frac{D_v^2(\vec{\phi})}{K_BT}\sum_\alpha c_\alpha\phi_\alpha \cdot eZ^*(\phi_i)\psi\right] - \frac{v}{n}(c_v - \sum_\alpha \phi_\alpha c_{eq}^\alpha)exp(\frac{-\sum_\alpha \phi_\alpha E_M^\alpha}{K_BT}) \quad (10)$$

where $M_{\alpha\beta}$ is the interfacial mobility, and $N_p$ serves as the number of coexisting phases in the neighboring grid points. $D$ is the interdiffusivity parameter as a function of the field order parameter. The applied electric field and the solid media are assumed to be quasi-static and isotropic, respectively. Consequently, the distribution of the electrostatic potential over the domain is calculated assuming the quasi-stationary conduction process. Hence, Ohm's equation (eqn. (8)) is used to obtain the electrostatic potential over the domain. $\kappa$ is the conductivity, and $\psi$ is the electrostatic potential. Equation of mechanical equilibrium (eqn. (9)) is solved iteratively to obtain the static equilibrium, the elastic energy is calculated subsequently. $\sigma_{ij}^{elas}$ is the elastic stress and $r_j$ is the j$^{th}$ component of position vector, $r$. Higher order solutions for the equation of mechanical equilibrium is derived using lower order approximations till satisfying the tolerance of $10^{-8}$ in displacements. The elastic strain (n$^{th}$ order approximation) is obtained using *Fourier* spectral approach [16,17]. Equation (10) is the vacancy diffusion equation taking into account the fluxes due to the difference in intrinsic diffusivity of elements (Cu and Sn) in different features of the microstructure, EM-mediated flux of vacancies in the reverse direction of atomic flux, and annihilation of vacancies. $c_{eq}^\alpha$ is the equilibrium concentration of vacancies in the phases, and $E_M^\alpha$ is the migration energy of vacancies.

The composition of coexisting phases at a given point is constrained by the condition of equality of chemical potential and mass conservation as:

$$f_{c_1}^1[c_1(x,t)] = f_{c_2}^2[c_2(x,t)] = \cdots = f_{c_N}^N[c_N(x,t)] \equiv f(x,t) \quad (11)$$

$$C = \sum_{\alpha=1}^N \phi_\alpha c_\alpha \quad (12)$$

where $c_N$ subscripts under the free energy densities denote the derivative of the free energy term with respect to concentration of the phase.

**Nucleation of Secondary Phases**

Classical nucleation theory is combined with a stochastic probabilistic framework to account for nucleation of IMC grains during the isothermal reactions between the Cu substrate and the Sn solder. The heterogeneous nucleation behavior of IMCs [18] is modeled by considering that nucleation is an stochastic event that can be modeled by means of a Poison distribution. In this respect, the nuclei of *Cu$_6$Sn$_5$* and *Cu$_3$Sn* IMCs are explicitly seeded in the microstructure by evaluating the probability of a nucleation event at any (untransformed) region where nucleation is likely to happen—i.e. at interfaces––at a specific time. The stochastic nucleation probability is approximated by the Poisson distribution:

$$P_n = 1 - \exp[-(J.v.\Delta t)] \quad (13)$$

Where $J$ is the nucleation rate, $v$ is volume of nucleus, and $\Delta t$ is the time interval of nucleation. Consequently, the nucleation rate can be obtained for an undercooled liquid systems [19]:

$$J = J_0 \exp\left[-\frac{16\pi\sigma^3}{3k_B T (\Delta G_v)^2}\left(\frac{Cos^3\theta - 3Cos\theta + 2}{4}\right)\right] \quad (14)$$

where $J_0$ is the nucleation rate constant utilized as a frequency factor. $T$ is the temperature, $k_B$ is the Boltzmann constant, $\sigma$ is the interface energy, $\Delta G_v$ is the driving force for nucleation and $\theta$ is the contact angle between nuclei and substrate. In the calculation of the stochastic probability of nucleation, we accounted for the variation in nucleation rate constant and the contact angle of the embryo with the substrate. The interval of nucleation, the time of seeding the nucleus into the domain, and other parameters in equation (13) and equation (14) are obtained by fitting the nucleation theory to the material conditions and relevant experimental data [5,19,20].

## Results and Discussion

**Microstructure generation**

Multi-phase-field modeling is used to study the formation of a Cu/Sn/Cu microjoint in the 3DIC packaging systems. Experimental studies show three different microstructural patterns for Cu/Sn/Cu microjoint depending on the processing conditions. These microstructural patterns are composed of Cu and Sn as initial materials, and $Cu_3Sn$ and $Cu_6Sn_5$ IMCs as the consequence of reaction between Cu and Sn. These patterns are reported in

Table 1. Accordingly, we first use multi-phase-field to study the formation of these patterns and then analyze the response of the microstructures under the accelerated EM conditions. This first step forms the necessary initial microstructures during virtual thermal treatment conditions, and helps to realize and understand the appropriate thermodynamic and kinetic conditions that are dominant locally during chemical reactions.

*Table 1 Microstructure case studies*

| Case | Microstructure pattern |
|---|---|
| Initial state | **Cu**/Sn/**Cu** |
| Case I | **Cu**/$Cu_3Sn$/$Cu_6Sn_5$/**Sn**/$Cu_6Sn_5$/$Cu_3Sn$/**Cu** |
| Case II | **Cu**/$Cu_3Sn$/$Cu_6Sn_5$/$Cu_3Sn$/**Cu** |
| Case III | **Cu**/$Cu_3Sn$/**Cu** |

The calculated microstructures are shown in Figure 2. In general, low temperature and low duration thermal treatments yield a microstructure where the solder material is still present in the interlayer (Refer to Figure 2(a)). This is also valid for larger soldering bumps such as C4 and BGAs. In this stage, the multi-phase-field model takes care of the solid and liquid state reactions at different locations of the simulation cell. Liquid state reaction proceeds until consumption of Sn layer entirely. Increasing thermal treatment temperature and/or duration make all solder material to be consumed by the $Cu_6Sn_5$ IMC layers. Hence, we may have microstructures that contain both $Cu_3Sn$ and $Cu_6Sn_5$ IMCs as shown in Figure 2(b). Furthermore, the growth regime changes in favor of $Cu_3Sn$ after all of the pure Sn phase is consumed. Accordingly, the kinetics of the reaction change, and $Cu_3Sn$ IMCs consume $Cu_6Sn_5$ IMCs at the expense of Sn and after a while $Cu_3Sn$ phases cover the entire interlayer. Another sequence of nucleation of $Cu_3Sn$ takes place at this state. Hence, the number density of $Cu_3Sn$ IMCs is higher at this stage. Such a microstructure is depicted in Figure 2(c). These three sets of microstructures are the consequence of dimensional limitations of microbumps in 3DIC systems, and also different/successive thermal treatment conditions that are usually utilized during reflowing process.

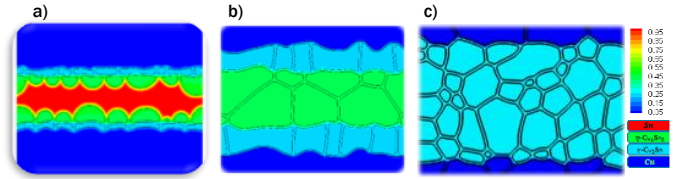

*Figure 2 a) Microstructure case I [3]: Sn-retained microstructure for the joint treated at low durations/low temperatures. b) Microstructure case II [3]: Sn-exhausted microstructure based on the second pattern in Table 1. c) Microstructure case III: $Cu_3Sn$ intermetallics covering the entire interlayer of the microstructure in LVIs for longer treatment durations/high temperatures.*

Figure 3 shows an example of the calculated polycrystalline microstructure with Cu/$Cu_3Sn$/Cu pattern in comparison with the real microstructure. The microstructure is in close agreement with the real case [21] which demonstrates the prosperous integration of 2 $\mu m$ TSV and Cu-Sn SLID interconnection between stacked chips.

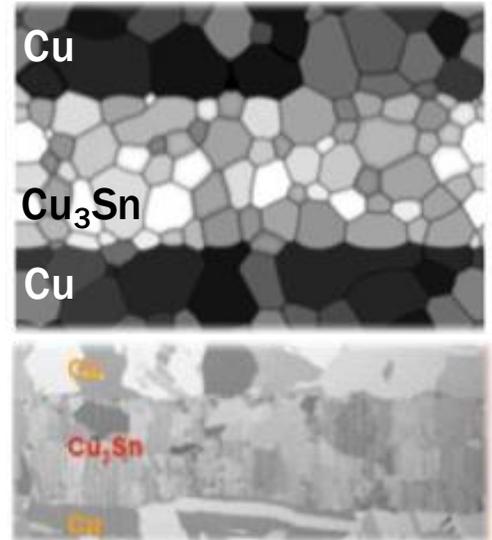

*Figure 3 The calculated microstructure (Case III) using multi-phase-field method versus experimental microstructure [21]*

**Electromigration response**

We have studied the EM response of these micro-joints under accelerated EM conditions by applying current densities in the range of $4\times 10^7$ at 180°C. Here we used equations 10, and 11 in a coupled way with the phase-field equations to perform elastic-electrical-chemical multi-phase-field calculations to study the evolution of diffusion-driven phenomena in the sandwich interconnection systems.

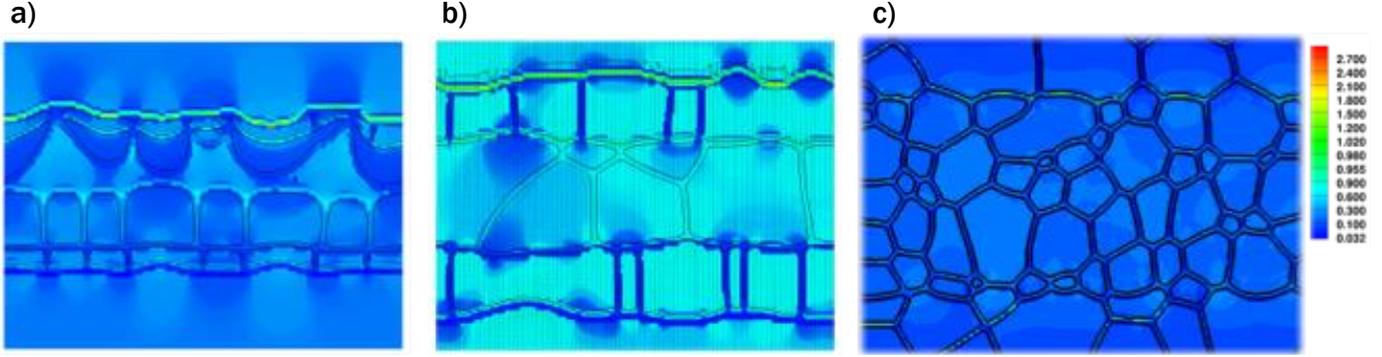

*Figure 4 Normalized current density distribution at 180°C under 4×10⁷ Am⁻² in the microstructures. a) case I b) case II, and c) case III. Less current divergences are present in microstructure case III.*

The calculations show that the wind force and subsequent current stressing effect enhance diffusion and alter reaction kinetics in the direction of electric field, resulting in a new dominant kinetic mechanism. In other words, EM changes the IMC growth trend in the multi-layered interconnection systems as observed in Refs. [6,12]. Our previous study [3] consistently showed this effect under $4\times10^7$ Am$^{-2}$ or higher current densities for microstructure case I, and under $4\times10^8$ Am$^{-2}$ or higher current densities for microstructure case II.

This study shows that microstructure case III has better response due to presence of a lot of vertical or semi-vertical fast diffusive channels in Cu$_3$Sn. In addition, since resistivity of Cu$_3$Sn (9.83) is relatively closer to the resistivity of Cu (1.70), less EM flux divergences happens in the microstructure. This has a direct impact on the convective flux of atoms and allows atoms to migrate easily without inducing further blocking. The snapshots of distribution of current densities in the microstructures under $4\times 10^7$ Am$^{-2}$ at 180°C are shown in Figure 4.

**Residual stresses due to lattice mismatches between the phases**

Microelasticity is the regime in which a material undergoes a phase transition along with elastic interactions due to misfit strains and inhomogeneity in elastic properties. Crystallographic misfits result in a volume change when a phase nucleation or transition happens. The stress-free strains that would result from unconstrained crystallographic misfits are eigenstrains, denoted $\varepsilon_{ij}^*$. The accommodation of these strains in elastically heterogeneous materials, e.g. multiphase and/or polycrystalline material systems determine the development of stress $\sigma_{ij}^{el}$ and elastic strain $\varepsilon_{ij}^{el}$ fields. By taking into account the appropriate mismatches and elastic inhomogeneities in elastic modulus of the phases, the elastic strain and stress distribution are calculated. Accordingly, the respective elastic driving force for evolution of the multi-phase microstructure is obtained as it is proposed in Ref. [22].

The snapshots of elastic strain and stress distribution in the simulation cell for microstructure case I are depicted in Figure 5. The elastic strain and stress, and the subsequent calculated elastic driving force determines whether a phase transition with expansion or contraction can be favored or hindered. In these calculations, Cu$_6$Sn$_5$ IMC layers are in direct contact with solid state Sn. The calculations show that Sn is under tension due to the growth of Cu$_6$Sn$_5$ from top and bottom side. Both Cu$_3$Sn and Cu$_6$Sn$_5$ IMCs are under compression. Anisotropy in the GB energy, the larger the misorientation angle between grains the faster the shrinkage rate and thicker the grain boundary.

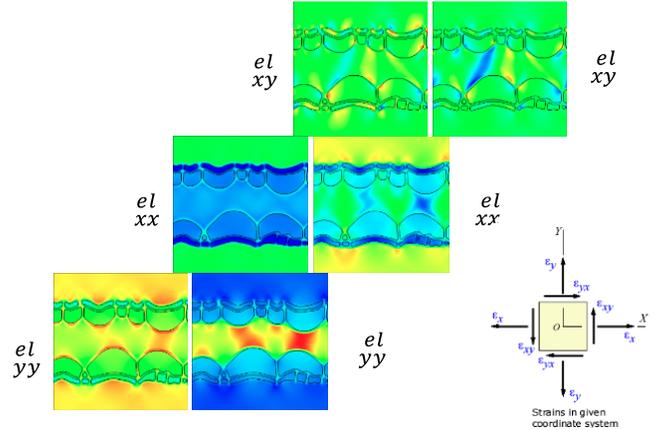

*Figure 5 Snapshots of elastic strain and stress distribution in the simulation cell for microstructure case I. Cu$_6$Sn$_5$ IMC layers are in direct contact with solid state Sn.*

**Supersaturation of non-equilibrium vacancies**

The vacancy evolution trend is calculated by coupling equation (10) with the phase-field equations and the result is shown in Figure 6. Vacancy flux is governed by the differences in intrinsic diffusivities of elements (Cu and Sn) in each feature of the microstructure, and the induced convective vacancy flux due to applied unidirectional electric field. The simulations show that vacancies evolve in the opposite direction of atomic flux. The results agree with literature where an increase in the vacancy amount in the *Cu$_3$Sn* layer at the cathode side is expected (red region in Fig. 6). On the other hand, the vacancy amount in the *Cu$_3$Sn* layer at the anode side decreases (blue region in the same figure). The change on the order of EM-mediated mass flux from TSV to the Sn phase (interlayer) is more than fourteen orders of magnitude in the Sn-retained structure. However, the flux only changes about eight orders of magnitude in the Sn-exhausted microstructure.

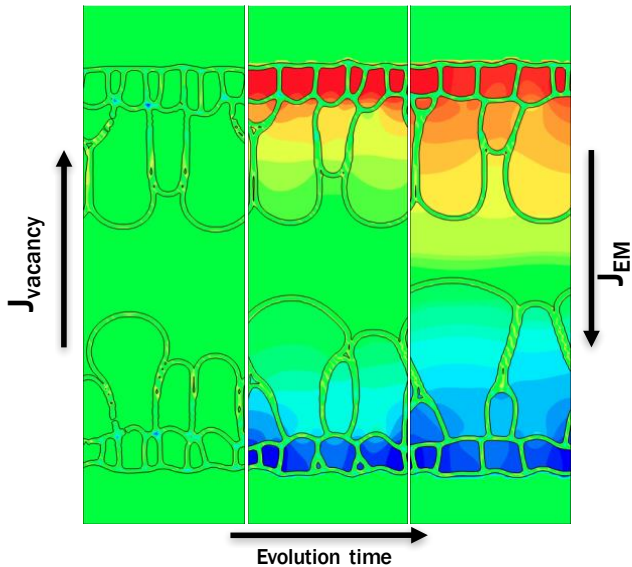

*Figure 6 Non-equilibrium vacancy evolution over time along with faster growth of IMCs at anode layer (bottom) under $4\times10^7\,Am^{-2}$ at $180°C$.*

The experimental observations show that the voids form after certain amount of time in the $Cu_3Sn$ IMCs of the cathode layer [7]. In this respect, the calculations suggest that the migration of atoms from cathode side to the anode side will leave vacancies behind. It is important to consider the rate of such a migration. Obviously, the rate is not consistent in the entire path and changes from one phase to the other one. In addition, the interfaces play different roles in this path, as they act as generators or annihilators of vacancies. While the experiments suggest that the short-circuit channels and their quantity play important role as vacancy sink locations in prevention of the formation of voids, the distinct role of each of these channels are rarely addressed in the literature. This study illustrates that the horizontal $Cu_6Sn_5/Cu_3Sn$ interface act as the vacancy generation source in the cathode layer, while the same interfaces in the anode side acts as a vacancy annihilation sinks. On the other hand, other vertical interfaces and the GBs serve as the rapid pathways for the migration of atoms along the direction of the applied current.

It is important to develop, and combine the existing model, with failure prediction models [23–26] to study the microstructural effects upon ductile failure in the LVI solder joints to effectively retain good strength and ductility during accelerated EM conditions in 3DIC systems.

## Conclusions

The multi-physics multi-phase-field approach is integrated and utilized to study EM induced mass transport and the subsequent impact on IMC growth. IMC growth is enhanced by electric current at the anode in the expense of IMCs at the cathode. It is shown that IMC covered microstructure has higher resistivity to EM induced severe mass fluxes.

## Acknowledgments

The authors would like to acknowledge the Terra supercomputing facility of Texas A&M University for providing computing resources useful in conducting the research reported in this paper. This research was supported by the National Science Foundation under NSF Grant No. CMMI-1462255.